\documentclass[conference]{IEEEtran}
\IEEEoverridecommandlockouts
\usepackage{cite}
\usepackage{amsmath,amssymb,amsfonts}
\usepackage{algorithmic}
\usepackage{graphicx}
\usepackage{textcomp}
\usepackage{xcolor}
\def\BibTeX{{\rm B\kern-.05em{\sc i\kern-.025em b}\kern-.08em
    T\kern-.1667em\lower.7ex\hbox{E}\kern-.125emX}}
\usepackage[utf8x]{inputenc}

\title{Study of Optical Networks, 5G, Artificial Intelligence and Their Applications 
}

\author{
\IEEEauthorblockN{Quanda Zhang}
\IEEEauthorblockA{School of Telecommunication\\
Daqing Normal College}
\and
\IEEEauthorblockN{Qi Zhang}
\IEEEauthorblockA{School of Telecommunication\\
Daqing Normal College}
}

\begin{document}

\maketitle

\begin{abstract}
This paper discusses the application of artificial intelligence (AI) technology in optical communication networks and 5G. It primarily introduces representative applications of AI technology and potential risks of AI technology failure caused by the openness of optical communication networks, and proposes some coping strategies, mainly including modeling AI systems through modularization and miniaturization, combining with traditional classical network modeling and planning methods, and improving the effectiveness and interpretability of AI technology. At the same time, it proposes response strategies based on network protection for the possible failure and attack of AI technology.
\end{abstract}


\section{Introduction}
Artificial intelligence (AI, artificial intelligence) technology is very early
has been used in many fields, but for many years this technology has not
gained high attention until AlphaGo defeated Chinese and Korean Go players
After the hand, it began to become a research hotspot, and researchers tried to
AI technology is applied in different fields, including optical communication network
network. In the past two years, the United States Optical Communications Conference (OFC, optical fiber
communication) and the European Conference of Optical Communications (ECOC, European conference of optical communication), at least 16 conference topics focused on AI or machine learning (ML,
machine learning) technology. This paper combines AI technology and ML technology
AI technologies are regarded as the same class of technologies, and at the same time, although AI technologies cover a wide range,
The AI technology referred to in this article is mainly neural network technology.

AI technology has received widespread attention mainly due to the following two reasons. First, AI technology is relatively easy to get started and use. it comes in black
Model the system in a box way, through a large number of samples
Learning, let the black box connect neurons by itself, and distribute neurons
Connect weights without requiring the user to understand why neurons behave the way they do
connections and are assigned current weights. Users only need to provide enough learning samples, increase the number of neurons and the number of hidden layers,
It can improve the prediction accuracy of AI technology. Second, AI technology is
After the AlphaGo incident, it has almost been deified, and almost everyone knows that "people
"artificial intelligence", and in the academic circle, the paper labeled AI
Papers also seem to be easier to publish, so this also leads to a current phenomenon
Like, that is, for almost all problems, regardless of whether it is suitable or not, the use of
AI technology for modeling and solving.

AI technology is very successful in solving some problems, as before
Go and some image-to-speech recognition scenarios mentioned above, but cannot
Due to the successful solution of a certain field or certain problems, AI is regarded as a "universal method". This paper aims at the current AI technology in optical communication
Discuss the application of AI technology in the network, including the application of AI technology in optical communication network
applicability in the network, and raises the potential risk of using AI technology
Some coping strategies.

\section{AI Applications in Optical Networks}

AI technology has been widely used in the literature of optical communication networks~\cite{zhang2014dynamic,lu2017game,jin2018multi,gong2013virtual,yin2013spectral,ciceri2021federated,chen2019deeprmsa}.
A great deal of research can be found in this area. This paper introduces several representative applications of AI technology
 in optical communication networks. 1) On receiving
At the end, using the digital signal processing method combined with AI technology, it can effectively improve the detection sensitivity of optical signals and improve the optical fiber transmission system
performance and improve the spectrum utilization efficiency of the network \cite{chen2019multi,chen2019deeprmsa}. 2) In the optical network, there are a large number of end-to-end optical channels, and these optical channels are respectively
related parameters (including transmission rate, modulation format,
number of optical fiber links, number of optical amplifiers and gain, etc.) and their receiving
the signal transmission quality (QoT, quality of transmission) detected by the end is used as input and output, and through a lot of learning, it can be realized
Prediction of QoT for different end-to-end optical channels in optical networks; where QoT
often expressed as the signal-to-noise ratio of the optical channel (OSNR, optical signal to
noise ratio), its accurate prediction can reduce the optical channel OSNR margin
configuration, thereby improving the spectrum utilization efficiency of the network \cite{chen2019demonstration,liu2018realizing}. 3) pass
Continuously learn the fault events in the optical network, and use the fault and fault cause
Because it is used as input and output, it can accurately analyze and diagnose the cause of the fault early warning of future failures ~\cite{jin2018multi}. 4) Combined with the need for network security, AI technology can also be used for early warning and identification of network attacks on the optical layer \cite{tian2021reconfiguring}.

\section{AI Applications in 5G Communication}

AI and ML are being utilized in 5G and mmWave communication~\cite{lu2017convolutional,shafique2020internet,alkhateeb2014channel,bhartia1984millimeter,rappaport2013millimeter,pi2011introduction,marcus2005millimeter} to enhance performance, reduce costs, and boost efficiency. Applications of AI in this field include network optimization, predictive maintenance, self-organizing networks, traffic prediction, security, resource allocation, network slicing, edge computing, interference management, and spectrum management. These technologies are still in the early stages of development but have the potential to significantly improve the performance, efficiency, and cost-effectiveness of these networks. As AI technologies continue to evolve and become more widely adopted, they will likely play an increasingly important role in the development and deployment of 5G and mmWave communication systems. However, it is also crucial to consider potential risks and challenges such as ensuring privacy, security, and ethical considerations.

AI techniques apply the same “black box” approach to
different application scenarios leading to method innovation and analysis of the underlying mechanism slack. A very typical example is as follows. Thanks to AI technologies such as deep learning can effectively recognize some image patterns, there are studies that the researchers applied this technology to the identification of lesions in different parts of the human body~\cite{shrikumar2016not,sendak2020human}.
Based on the same method and process, different human parts are continuously used bit pictures, which can form a large number of so-called "research results" and thesis. Obviously, from the perspective of cultivating students and scientific research, students
The development of research skills and professionalism acquired practically in the project
are very few, and the actual work is only to collect relevant image data and
Write a small amount of Python code, and finally hand over the training task to the graph
Processor (GPU, graphics processing unit) to complete, no
Carry out in-depth thinking on the method and mechanism of specific research questions
It is obviously not conducive to innovation and effective innovation, and it is impossible to grasp (in fact, it is currently impossible to grasp) what is going on in the black box.

\bibliographystyle{IEEEtran}
\bibliography{reference}

\end{document}